# ZOS: A Fast Rendezvous Algorithm Based on Set of Available Channels for Cognitive Radios

Zhiyong Lin, Lu Yu, Hai Liu, Yiu-Wing Leung, and Xiaowen Chu

*Abstract*— Most of existing rendezvous algorithms generate channel-hopping sequences based on the whole channel set. They are inefficient when the set of available channels is a small subset of the whole channel set. We propose a new algorithm called ZOS which uses three types of elementary sequences (namely, Zero-type, One-type, and S-type) to generate channel-hopping sequences based on the set of available channels. ZOS provides guaranteed rendezvous without any additional requirements. The maximum time-to-rendezvous of ZOS is upper-bounded by $O(m_1 \times m_2 \times \log_2 M)$ where $M$ is the number of all channels and $m_1$ and $m_2$ are the numbers of available channels of two users.

*Index Terms*—Cognitive radio, rendezvous, channel hopping.

## I. INTRODUCTION

COGNITIVE radio (CR) networks can improve spectrum utilization in a way that CRs sense and opportunistically access temporarily idle channels licensed to the primary users (PUs). To communicate with each other, two users (CRs) should meet and establish a communication link on a commonly available channel. This process is called *rendezvous* [1]. Channel-hopping (CH) technique is a representative approach to rendezvous. With the CH technique, the network is time-slotted and each user hops on one channel in each timeslot according to a certain CH sequence. Two users are said to *achieve rendezvous* if they hop on the same available channel in the same timeslot. The number of timeslots that it takes to achieve rendezvous is defined as *time-to-rendezvous* (TTR). A CH algorithm is said to provide *guaranteed rendezvous* if its maximum TTR (MTTR) is finite.

A number of CH algorithms have been proposed in the literature and some of them (e.g., EJS [2], SSB [3], and E-AHW [4]) provide guaranteed rendezvous. These algorithms generate the CH sequences based on the whole channel set (i.e., the set of all potentially available channels) and their MTTRs are upper-bounded by expressions of the cardinality of the whole channel set. However, a user usually identifies a small portion of the whole channel set as its available channels in practice. For example, work in [5] shows that at most 15% of channels can be correctly detected as "available" when the signal-to-noise-ratio is −3dB. If a user follows a CH sequence which is designed based on the whole channel set, it may take a long time to achieve rendezvous with its potential neighbors. Thus, it is more desirable to design CH algorithms based on the set of available channels. There are two existing algorithms, i.e., Modified Modular Clock (MMC) [1] and CSAC [6], which generate CH sequences based on the set of available channels. However, it is already known that MMC cannot provide guaranteed rendezvous [1] and two users are required to play sender and receiver roles, respectively, in CSAC [6].

In this work, we propose a new CH algorithm called ZOS which is named after the three types of elementary sequences used in the algorithm, i.e., Zero-type, One-type, and S-type. ZOS advances the state-of-the-art in the following sense: 1) ZOS generates CH sequences based on the set of available channels and provides guaranteed rendezvous without any additional requirements; 2) ZOS gives a lower upper-bound of MTTR which is an expression of the cardinality of the set of available channels (see Theorem 2 in Section III). Table I summarizes the characteristics and performance of different algorithms.

TABLE I. COMPARISON OF DIFFERENT ALGORITHMS

| Algorithms | Additional requirements | Based on the set of available channels | Guaranteed rendezvous | Upper-bound of MTTR |
|---|---|---|---|---|
| **ZOS** | **Null** | **YES** | **YES** | $(12\lceil \log_2 M \rceil + 2)(P_1 P_2 + \max\{P_1, P_2\})$ |
| MMC [1] | Null | YES | NO | Infinite |
| CSAC [6] | Sender/receiver role | YES | YES | $O(m_1 P_2)$, when $m_1$ is divisible by $P_2$; $O(m_1 m_1 P_2)$, otherwise |
| SSB [3] | Null | NO | YES | $(M-1)(2M-1)$ |
| EJS [2] | Null | NO | YES | $4P(P+1-G)$ |
| E-AHW [4] | User ID | NO | YES | $147P(M+1-G)$ |

Remarks: 1) We focus on the asymmetric model where different users may have different available channels. The asymmetric model is a general case of the symmetric model where all users have same available channels [2] [3] [4]. 2) $M$ is the number of all potentially available channels; $P$ is the smallest prime number such that $P>M$; $m_1$ and $m_2$ are the numbers of channels available to user 1 and user 2, respectively; $P_1$ and $P_2$ are the smallest prime numbers such that $P_1 \geq m_1$ and $P_2 \geq m_2$, respectively; $G$ is the number of channels commonly available to the two users.

## II. ALGORITHM

We focus on the rendezvous between two users, say, users 1 and 2. We use $M$ ($M>1$) to denote the size of whole channel set $C=\{c_1, c_2, \ldots, c_M\}$, where $c_i$ ($i=1, 2, \ldots, M$) denotes the $i^{th}$ channel (channel

$i$). Each user is equipped with one CR which can access the idle channels in $C$ opportunistically. Let $C_1 \subseteq C$ and $C_2 \subseteq C$ denote the sets of channels available to users 1 and 2, respectively, where a channel is available to a user if the user can communicate on the channel without causing interference to the PUs. We assume that the PUs will not appear in the available channels of the users during the rendezvous operation. Note that $C_1$ is usually different from $C_2$ due to different locations of the users and their possibly heterogeneous capabilities in spectrum sensing. We assume there is at least one channel commonly available to users 1 and 2, i.e., $C_1 \cap C_2 \neq \phi$, which is a necessary condition of rendezvous.

*A. Algorithm description*

We use $X=<X[1], X[2], \ldots, X[n]>_{1 \times n}$ to denote a sequence consisting of $n$ items, and $X[i:j]=<X[i], X[i+1], \ldots, X[j]>_{1 \times (j+1-i)}$ to denote a subsequence of $X$ ($1 \leq i \leq j \leq n$). Given sequences $X=<X[1], X[2], \ldots, X[n]>_{1 \times n}$ and $Y=<Y[1], Y[2], \ldots, Y[m]>_{1 \times m}$, $<X, Y>$ denotes the resulting sequence by joining $X$ and $Y$, i.e., $<X, Y>=<X[1], X[2], \ldots, X[n], Y[1], Y[2], \ldots, Y[m]>_{1 \times (n+m)}$.

Our basic idea is as follows. According to our preliminary work [6], if two users play distinct roles (namely, sender and receiver roles) and use distinct sequences, rendezvous can be efficiently achieved. However, in some applications such as neighbor discovery, it may not be possible to pre-assign distinct roles to the users in advance. The goal of ZOS is to keep using distinct sequences for fast rendezvous while avoiding the assumption of pre-assigned sender/receiver role. To this end, each user selects one of its available channels and generates a "seed" (i.e., a binary sequence) based on binary representation of this channel index. If the two users select different channels, they generate distinct seeds (i.e., distinct binary sequences) and then they use the two respective types of sequences (i.e., the 0-type elementary sequence and the 1-type elementary sequence in ZOS) for fast rendezvous. If the two users select the same channel and generate the same seed, they can simply achieve rendezvous on this channel by letting the users stay on this channel (i.e., the s-type elementary sequence in ZOS).

We formally present the pseudo code of ZOS in Fig. 1, where each user selects a channel $s$ (called *stay channel*) from its available channel set in line 3 and generates its seed in line 6. The seed is a

$6L$-bit binary sequence plus one special bit (i.e., $6L+1$ bits in total, where $L=\lceil \log_2 M \rceil$). If the $i^{th}$ ($1\leq i\leq 6L$) bit of the seed is 0/1, it determines the 0-type/1-type elementary sequence. The last bit of the seed (i.e., the $(6L+1)^{th}$ bit) determines the s-type elementary sequence. The 0-type and 1-type elementary sequences are computed in line 8 where $Z^{(i)}$ denotes the elementary sequence determined by the $i^{th}$ ($1\leq i\leq 6L$) bit. The s-type elementary sequence consists of only one channel $s$ (line 9). As illustrated in Fig. 5(a), all the elementary sequences determined by the seed can form a two-dimensional table, where the length of the row equals the length of the seed and each column repeats the elementary sequence determined by the corresponding bit in the seed (i.e., the $i^{th}$ column repeats the $i^{th}$ elementary sequence $Z^{(i)}$). The user continuously performs channel hopping row-by-row (i.e., round-by-round) in the table until it achieves rendezvous with its neighbor (lines 11-15). Line 12 is to determine the $i^{th}$ elementary sequence $Z^{(i)}$ corresponding to timeslot $t$. Line 13 is to determine which item (i.e., channel) the user should access in timeslot $t$.

```
                    Algorithm ZOS
1: Input: M, available channel set c̃   //c̃ is either C₁ or C₂
2: L=⌈log₂M⌉;
3: s=the index of a channel randomly selected from c̃;
4: A=L-bit binary representation of s;
5: O=<0, 0, …, 0>₁ₓL,  I=<1, 1, …, 1>₁ₓL;
6: D=<A, O, I, A, O, I, s>₁ₓ(6L+1);  //D is the seed
7: for i=1, 2, …, 6L+1  //generate elementary sequences Z^(i)
8:    if (i≤6L)  Z^(i)=ZeroOneES(c̃, D[i]); //Z^(i) is 0-type or 1-type
9:    else       Z^(i)=<s>;  //Z^(6L+1) is s-type
10: t=1;
11: while (not rendezvous)
12:    i=((t−1) mod (6L+1))+1;  //each round consists of 6L+1 timeslots
13:    n=((⌈t/(6L+1)⌉−1) mod |Z^(i)|)+1;  //|Z^(i)| denotes the length of Z^(i)
14:    Attempt rendezvous on channel Z^(i)[n];  //Z^(i)[n] is the n^th item of Z^(i)
15:    t=t+1;
```

Fig. 1. Pseudo code of the ZOS algorithm.

The 0-type and 1-type elementary sequences are generated by ZeroOneES function in Fig. 2. There are two steps in ZeroOneES. In the first step, a smallest prime number $P$ is selected such that $P\geq|\tilde{c}|$. Then, as described in lines 4-11 of Fig. 2, two sequences $X$ and $Y$ are constructed such that $X$ consists of $P$ items and $Y$ consists of $P+b$ items (i.e., $P+1$ items when $b=1$ and $P$ items when $b=0$). In line 5 of Fig. 2, $per(\tilde{c})$ generates a random permutation of channel indices in $\tilde{c}$, e.g., $per(\tilde{c})$ may output <3, 1, 4> if $\tilde{c}=\{c_1, c_3, c_4\}$. In line 7 of Fig. 2, $select(\tilde{c}, k)$ generates a sequence consisting of $k$ channel indices randomly selected from $\tilde{C}$, e.g., $select(\tilde{C}, 3)$ may output <1, 3, 1> if $\tilde{C}=\{c_1, c_3, c_4\}$. In the second step, the resulting sequence $Z$ is constructed by properly interleaving sequences $X$ and $Y$ as follows: 1) If $b=1$, $Z$ has $2P(P+1)$ items. The items with odd subscripts are obtained by repeating $X_{1\times P}$ $P+1$ times while the items with even subscripts are obtained by repeating $Y_{1\times(P+1)}$ $P$ times; 2) If $b=0$, $Z$ has $2P$ items with $<Z[1], Z[3], …, Z[2P−1]>_{1\times P}=X_{1\times P}$ and $<Z[2], Z[4], …, Z[2P]>_{1\times P}=Y_{1\times P}$.

```
Function ZeroOneES
 1: Input: available channel set C̃, binary value b (0 or 1)
 2: Output: Z, which is a sequence consisting of channel indices in C̃
 3: m=|C̃|, P=the smallest prime number that is not less than m;
 4: X=<0, 0, …, 0>_{1×P};
 5: X[1: m]=per(C̃);  //the first m items of X form a permutation of C̃
 6: if (P>m)  //set the remaining P−m items of X
 7:     X[m+1: P]=select(C̃, P−m);
 8: Y=<0, 0, …, 0>_{1×(P+b)};
 9: Y[1: m]=per(C̃);  //the first m items of Y form a permutation of C̃
10: if ((P+b)>m)  //set the remaining P+b−m items of Y
11:     Y[m+1: P+b]=select(C̃, P+b−m);
12: if (b= =0)  //generate 0-type elementary sequence
13:     for i=1, 2, …, 2P
14:         if ((i mod 2)= =1)      Z[i]=X[(i+1)/2];
15:         else                    Z[i]=Y[i/2];
16: else  //b=1, generate 1-type elementary sequence
17:     for i=1, 2, …, 2P(P+1)
18:         if ((i mod 2)= =1)
19:             j=(((i+1)/2−1) mod P)+1,  Z[i]=X[j];
20:         else
21:             j=((i/2−1) mod (P+1))+1,  Z[i]=Y[j];
22: return Z;
```

Fig. 2. Pseudo code of the ZeroOneES function.

*B. Illustration examples*

We suppose that $C=\{c_1, c_2, c_3\}$ (i.e., $M=3$), $C_1=\{c_1, c_2\}$ and $C_2=\{c_2, c_3\}$. Note that channel 2 is the only channel commonly available to both users 1 and 2. Fig. 3 shows sample 0-type and 1-type elementary sequences generated by ZeroOneES.

Fig. 4 demonstrates that two users achieve rendezvous as long as one user performs the 0-type elementary sequence and the other performs the 1-type elementary sequence.

Fig. 5 shows the sample sequences of users 1 and 2 performing the ZOS algorithm. For simplicity, we let user 1 adopt the 0-type/1-type elementary sequence in Fig. 3(a)/3(b), and let user 2 adopt the 0-type/1-type elementary sequence in Fig. 3(c)/3(d). We point out that the 0-type/1-type elementary sequences could be different in a CH sequence of a user since ZeroOneES is called $6L$ times (line 8 of Fig. 1) and a random permutation is generated each time (lines 5 and 9 of Fig. 2).

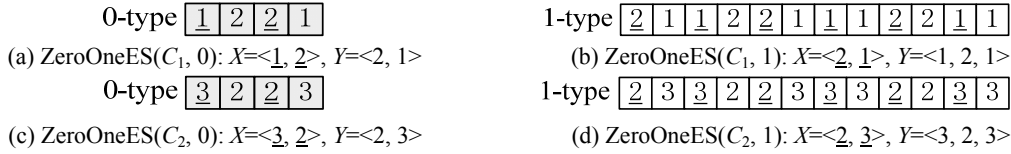

(a) ZeroOneES($C_1$, 0): $X$=<$\underline{1}$, 2>, $Y$=<2, $\underline{1}$>  
(b) ZeroOneES($C_1$, 1): $X$=<$\underline{2}$, $\underline{1}$>, $Y$=<1, 2, $\underline{1}$>  
(c) ZeroOneES($C_2$, 0): $X$=<$\underline{3}$, 2>, $Y$=<2, $\underline{3}$>  
(d) ZeroOneES($C_2$, 1): $X$=<$\underline{2}$, $\underline{3}$>, $Y$=<3, 2, $\underline{3}$>  

Fig. 3. Sample 0-type and 1-type elementary sequences. The items with underlines are with odd subscripts and they are obtained by repeating $X$.

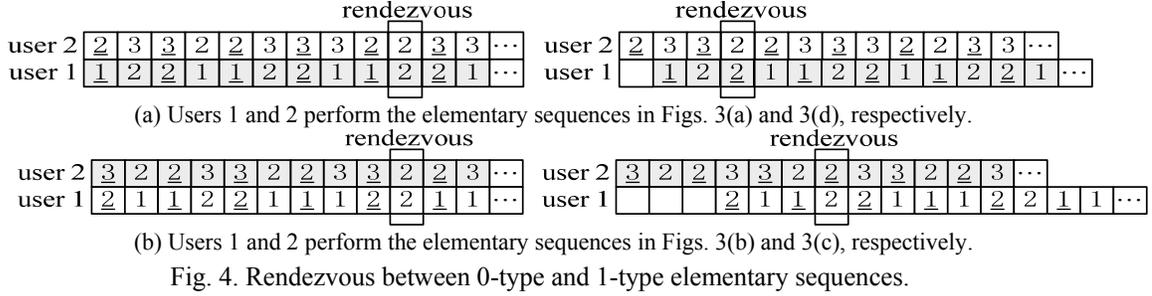

(a) Users 1 and 2 perform the elementary sequences in Figs. 3(a) and 3(d), respectively.

(b) Users 1 and 2 perform the elementary sequences in Figs. 3(b) and 3(c), respectively.

Fig. 4. Rendezvous between 0-type and 1-type elementary sequences.

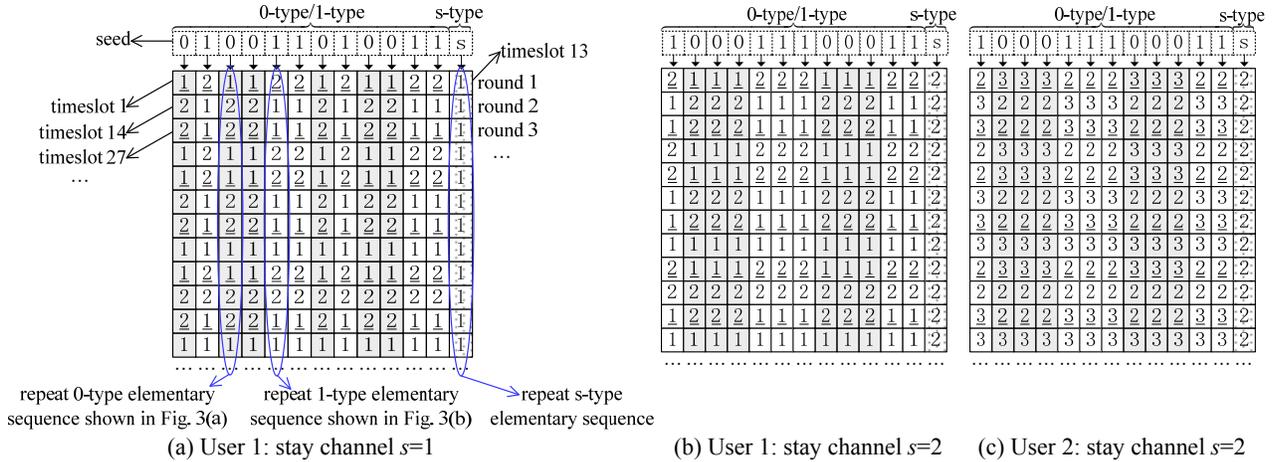

(a) User 1: stay channel $s$=1    (b) User 1: stay channel $s$=2    (c) User 2: stay channel $s$=2

Fig. 5. Sample sequences of users 1 and 2 performing ZOS. Each round consists of $(6L+1)=13$ timeslots where $M$=3 and $L=\lceil \log_2 M \rceil=2$.

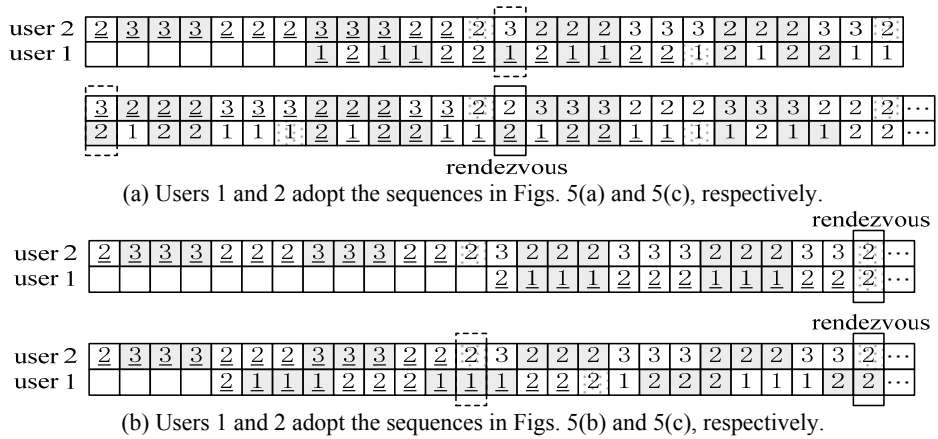

(a) Users 1 and 2 adopt the sequences in Figs. 5(a) and 5(c), respectively.

(b) Users 1 and 2 adopt the sequences in Figs. 5(b) and 5(c), respectively.

Fig. 6. Rendezvous is guaranteed using the sequences in Fig. 5. The blocks with solid lines indicate success of rendezvous while those with dashed lines do not.

With the CH sequences in Fig. 5, users 1 and 2 can achieve rendezvous under various scenarios, as illustrated in Fig. 6. In Fig. 6(a), the two users use the different stay channels. In this case, there is guaranteed rendezvous between the 1-type elementary sequence of one user and the 0-type elementary sequence of the other user. In Fig. 6(b), the two users use the same stay channel. In this case, there is guaranteed rendezvous between the s-type elementary sequence of one user (i.e., the stay channel) and the sequence of the other user.

## III. PERFORMANCE EVALUATION

In this section, we evaluate the performance of our ZOS algorithm via theoretical analysis and experimental study. Theorem 1 states that two users achieve rendezvous as long as they respectively perform the 0-type and the 1-type elementary sequences. Based on Theorem 1, Theorem 2 gives an upper-bound of the MTTR of ZOS.

**Theorem 1**. *Let $W^{(1)}=<Z^{(1)}, Z^{(1)}, Z^{(1)}, …>$ and $W^{(2)}=<Z^{(2)}, Z^{(2)}, Z^{(2)}, …>$, where sequence $Z^{(1)}$ is generated by ZeroOneES($C_1$, 1) and sequence $Z^{(2)}$ by ZeroOneES($C_2$, 0). If users 1 and 2 perform sequences $W^{(1)}$ and $W^{(2)}$ respectively, they achieve rendezvous in at most $2(P_1+1)P_2$ timeslots, where $P_1$ (resp., $P_2$) denotes the smallest prime number which is not less than $|C_1|$ (resp., $|C_2|$).*

*Proof:* Without loss of generality, assume user 2 is in timeslot $T$ ($1 \leq T$) with respect to its clock when user 1 starts its sequence, i.e., $W^{(1)}[t]$ is *aligned* with $W^{(2)}[T+t-1]$ ($1 \leq t$). We suppose that $T$ is an odd number (The case of even $T$ can be analyzed in a similar way). Let $X^{(1)}=<W^{(1)}[1], W^{(1)}[3], …>$, $Y^{(1)}=<W^{(1)}[2], W^{(1)}[4], …>$, $X^{(2)}=<W^{(2)}[T], W^{(2)}[T+2], …>$, and $Y^{(2)}=<W^{(2)}[T+1], W^{(2)}[T+3], …>$. It suffices to show there exist $c \in C_1 \cap C_2$ and $t$ ($1 \leq t \leq (P_1+1)P_2$) such that $X^{(1)}[t]=X^{(2)}[t]=c$ or $Y^{(1)}[t]=Y^{(2)}[t]=c$.

Note that $Z^{(1)}$ is generated by ZeroOneES($C_1$, 1). According to ZeroOneES we know: 1) $X^{(1)}$ and $Y^{(1)}$ are periodical and repeat $X^{(1)}[1: P_1]$ and $Y^{(1)}[1: P_1+1]$ respectively; 2) Every channel $i \in C_1$ is contained in both $X^{(1)}[1: P_1]$ and $Y^{(1)}[1: P_1+1]$. Similarly, since $Z^{(2)}$ is generated by ZeroOneES($C_2$, 0), $X^{(2)}$ and $Y^{(2)}$ repeat $X^{(2)}[1: P_2]$ and $Y^{(2)}[1: P_2]$ respectively, and every channel $j \in C_2$ is contained in both $X^{(2)}[1: P_2]$ and $Y^{(2)}[1: P_2]$.

We next show the following fact: *suppose sequences U and V are constructed by respectively*

repeating $\langle u_1, u_2, \ldots, u_p \rangle$ $q$ times and $\langle v_1, v_2, \ldots, v_q \rangle$ $p$ times, if $p$ and $q$ are co-prime, for any $i$ $(1 \leq i \leq p)$ and $j$ $(1 \leq j \leq q)$ there exists $k$ $(1 \leq k \leq pq)$ such that $U[k]=u_i$ and $V[k]=v_j$. Actually, since $p$ and $q$ are co-prime, according to the Chinese Remainder Theorem, given $i$ and $j$ we can find $k$ $(1 \leq k \leq pq)$ such that $(k-1) \equiv (i-1) \bmod p$ and $(k-1) \equiv (j-1) \bmod q$, i.e., we have $i=((k-1) \bmod p)+1$ and $j=((k-1) \bmod q)+1$. Since $U[h]=U[((h-1) \bmod p)+1]$ and $V[h]=V[((h-1) \bmod q)+1]$ $(1 \leq h \leq pq)$, we have $U[k]=U[((k-1) \bmod p)+1]=U[i]=u_i$ and $V[k]=V[((k-1) \bmod q)+1]=V[j]=v_j$.

Since $P_1$ and $P_2$ are prime numbers, $P_2$ must be co-prime either to $P_1$ or $(P_1+1)$. If $P_1$ and $P_2$ are co-prime, according to the above fact, for any $i \in C_1$ and $j \in C_2$ there exists $t$ $(1 \leq t \leq P_1 P_2)$ such that $X^{(1)}[t]=i$ and $X^{(2)}[t]=j$, meaning that for any $c \in C_1 \cap C_2$ there exists $t$ $(1 \leq t \leq P_1 P_2)$ such that $X^{(1)}[t]=X^{(2)}[t]=c$. If $P_1+1$ and $P_2$ are co-prime, we can similarly assert that for any $c \in C_1 \cap C_2$ there exists $t$ $(1 \leq t \leq (P_1+1)P_2)$ such that $Y^{(1)}[t]=Y^{(2)}[t]=c$.  ∎

**Theorem 2**. *If users 1 and 2 perform ZOS, the MTTR is upper-bounded by $(12\lceil \log_2 M \rceil+2)(P_1 P_2 + \max\{P_1, P_2\})$, which is not greater than $(24\lceil \log_2 M \rceil+4)(2|C_1||C_2|+\max\{|C_1|, |C_2|\})$, where $M$ is the number of all potentially available channels, and $P_1$ (resp., $P_2$) denotes the smallest prime number which is not less than $|C_1|$ (resp., $|C_2|$).*

*Proof:* Without loss of generality, assume user 2 is in its $n^{\text{th}}$ $(1 \leq n)$ round when user 1 starts performing ZOS. According to ZOS, each user should select its stay channel (see line 3 of Fig. 1). Let $s_1$ and $s_2$ be the stay channels of users 1 and 2, respectively. There are two cases.

Case 1: $s_1 \neq s_2$. As shown in Fig. 7(a), the overlap between the first $6L$ timeslots in user 1's first round and the first $6L$ timeslots in user 2's $n^{\text{th}}$ or $(n+1)^{\text{th}}$ round must be not less than $3L$ (i.e., $3L \leq h$, see Fig. 7(a)). Let $D^{(1)}$ and $D^{(2)}$ be the corresponding seeds of users 1 and 2, respectively (see line 6 of Fig. 1). Since $s_1 \neq s_2$, we can easily verify $D^{(1)}[i: i+3L-1] \neq D^{(2)}[j: j+3L-1]$ for any $i$ and $j$ $(1 \leq i, j \leq 3L)$, that is, there exists $k$ $(0 \leq k \leq 3L-1)$ such that $D^{(1)}[i+k] \neq D^{(2)}[j+k]$ (One item is 1 and the other is 0). According to this fact, in the overlap there exists a position where one user is with 1-type and the other is with 0-type. Assume this position is corresponding to global timeslot $t_0$ (see Fig. 7(a)). According to ZOS, in global timeslots $\{t_0, t_0+6L+1, t_0+2(6L+1), \ldots\}$ one user should repeat its 1-type elementary sequence and the other should repeat its 0-type elementary sequence. By Theorem 1 we know that, if users 1 and 2 repeat the 1-type (reps., 0-type) and the 0-type (reps., 1-type) elementary sequences respectively, they rendezvous within $\tau_1=[2(P_1+1)P_2](6L+1)$ (resp., $\tau_2=[2(P_2+1)P_1](6L+1)$) timeslots. Thus, the TTR in Case 1 is not greater than $\max\{\tau_1, \tau_2\}=(12L+2)(P_1P_2+\max\{P_1, P_2\})$.

Case 2: $s_1=s_2=s$. As shown in Fig. 7(b), if user 1's first round is exactly aligned with user 2's $n^{th}$ round, they achieve rendezvous on channel $s$ with TTR equal to $6L+1$. Otherwise, the 0-type/1-type part in user 1's first round should overlap with the s-type part in user 2's $n^{th}$ round. Assume this overlap position is global timeslot $t_1$ (see Fig. 7(b)). According to ZOS, in global timeslots $\{t_1, t_1+6L+1, t_1+2(6L+1), \ldots\}$ user 1 should repeat the 0-type/1-type elementary sequence and must visit channel $s$ in some timeslot of $\{t_1, t_1+6L+1, t_1+2(6L+1), \ldots, t_1+(2P_1-1)(6L+1)\}$. This implies that the two users achieve rendezvous within $2P_1(6L+1)$ timeslots. Thus, the TTR in Case 2 is not greater than $\max\{6L+1, 2P_1(6L+1)\}=2P_1(6L+1)$.

To sum up, users 1 and 2 are guaranteed to rendezvous within $\tau=\max\{(12L+2)(P_1P_2+\max\{P_1, P_2\}), 2P_1(6L+1)\}= (12L+2)(P_1P_2+\max\{P_1, P_2\})$ timeslots. Since $P_1$ (resp., $P_2$) is the smallest prime number which is not less than $|C_1|$ (resp., $|C_2|$), according to the Bertrand-Chebyshev Theorem, we know that $P_1 \leq 2|C_1|$ (resp., $P_2 \leq 2|C_2|$). Hence, $\tau$ can be further upper-bounded by $(24L+4)(2|C_1||C_2|+\max\{|C_1|, |C_2|\})$. ∎

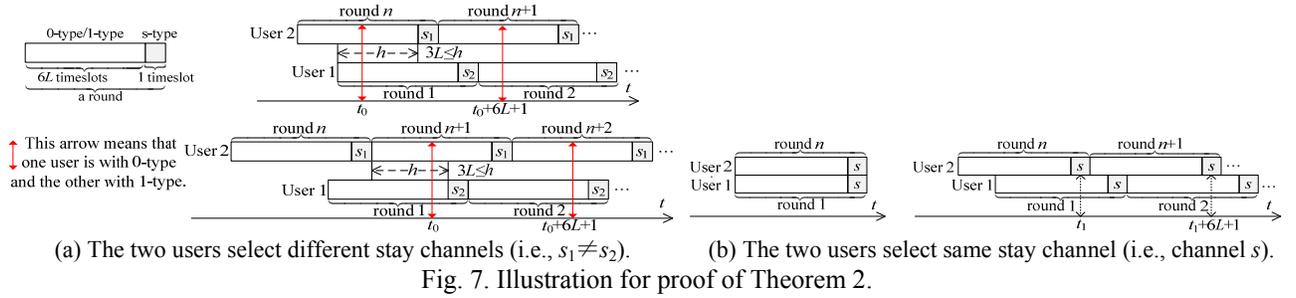

(a) The two users select different stay channels (i.e., $s_1 \neq s_2$).      (b) The two users select same stay channel (i.e., channel $s$).

Fig. 7. Illustration for proof of Theorem 2.

We conduct simulation in Matlab 7.11. CSAC [6], MMC [1], EJS [2], and SSB [3] are selected for comparison. E-AHW [4] is not included since it requires information of user ID and generates CH sequences based on the whole channel set. In the simulation, the size of $C$ is set to be 100. The available channels of each user are randomly selected from $C$ and each user has $\theta \times |C|$ ($0 \leq \theta \leq 1$) available channels on average, where $\theta$ varies from 0.1 to 0.5. The number of commonly available channels (i.e., $G$) is fixed at six. We report in Fig. 8 the simulation results on the average TTR and the maximum TTR. Each result is obtained from 5000 separate runs. We can see that ZOS significantly outperforms the existing algorithms in both the average TTR and the maximum TTR.

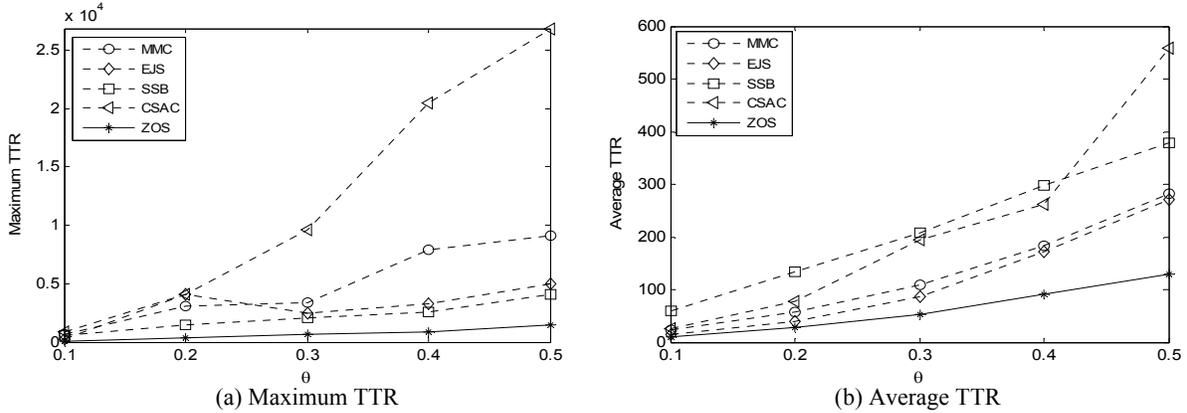

Fig. 8. Maximum TTR and average TTR of different algorithms.

## IV. Conclusion and Future Work

In this work, we focus on two-user rendezvous and present a new algorithm called ZOS which generates CH sequences based on the set of available channels. One possible multi-user rendezvous method is to iteratively perform two-user rendezvous as follows [7]. Once two users achieve rendezvous, they can exchange their available channel sets. In the next iteration, they perform two-user rendezvous with another user in a similar manner. This process is continued until every user obtains the available channel sets of all the other users. Then the users can select a channel from the intersection of the available channel sets of all users and use this channel for communication among themselves. In our future work, we will investigate more efficient multi-user rendezvous methods.